\documentclass[12pt,preprint]{aastex}
\pdfoutput=1

\shorttitle{DD Merger SNe}
\shortauthors{Fryer et al.}

\def \nuc#1#2{\relax\ifmmode{}^{#1}{\protect\text{#2}}\else${}^{#1}$#2\fi}

\begin{document}

\title{Spectra of Type Ia Supernovae from Double Degenerate Mergers}

\author{Chris L. Fryer\altaffilmark{1,2}, Ashley
  J. Ruiter\altaffilmark{3}, Krzysztof
  Belczynski\altaffilmark{1,4}, Peter J. Brown\altaffilmark{5},
  Filomena Bufano\altaffilmark{6}, Steven Diehl\altaffilmark{1},
  Christopher J. Fontes\altaffilmark{1}, Lucille
  H. Frey\altaffilmark{1}, Stephen T. Holland\altaffilmark{7}, Aimee
  L. Hungerford\altaffilmark{1}, Stefan Immler\altaffilmark{7}, Paolo
  Mazzali\altaffilmark{3,8,9}, Casey Meakin\altaffilmark{10}, Peter
  A. Milne\altaffilmark{10}, Cody Raskin\altaffilmark{11}, Francis
  X. Timmes\altaffilmark{11}}

\altaffiltext{1}{Los Alamos National Laboratory, Los Alamos, NM
87545}

\altaffiltext{2}{Physics Dept., University of Arizona, Tucson AZ
85721}

\altaffiltext{3}{Max-Planck-Institut f\"ur Astrophysik,
Karl-Schwarzschild-Str. 1, 85741 Garching, Germany}

\altaffiltext{4}{Astronomy Dept., New Mexico State University, Las
Cruces, NM 88003}

\altaffiltext{5}{Pennsylvania State University, Dept. of Astronomy 
\& Astrophsycis, University Park, PA 16802}

\altaffiltext{6}{Dipartimento di Astronomia, Univ. Padova,
  INAF-Osservatoroio Astronomico di Padova}

\altaffiltext{7}{Astrophysics Science Division, NASA Goddard Space 
Flight Center, Greenbelt, MD 20771}

\altaffiltext{8}{INAF-OAPD, vicolo dell'Osservatorio, 5, 35122, 
Padova, Italy}

\altaffiltext{9}{Scuola Normale Superiore, Piazza dei Cavalieri, 7, 56126 Pisa, Italy}

\altaffiltext{10}{Steward Observatory, 933 North Cherry Avenue, RM
N204, Tucson, AZ 85721}

\altaffiltext{11}{SESE, Arizona State University, Tempe, AZ 85287}

\email{fryer@lanl.gov, kbelczyn@nmsu.edu, grbpeter@yahoo.com,
  filomena.bufano@oapd.inaf.it, stevendiehl@gmail.com, lfrey@lanl.gov,
  cjf@lanl.gov, Stephen.T.Holland@nasa.gov, aimee@lanl.gov,
  stefan.m.immler@nasa.gov,
  mazzali@MPA-Garching.MPG.DE,pmilne511@cox.net,codyraskin@gmail.com,
  ajr@mpa-garching.mpg.de}

\begin{abstract}

The merger of two white dwarfs (a.k.a. double degenerate merger) has
often been cited as a potential progenitor of type Ia supernovae.
Here we combine population synthesis, merger and explosion models with
radiation-hydrodynamics light-curve models to study the implications
of such a progenitor scenario on the observed type Ia supernova
population.  Our standard model, assuming double degenerate mergers do
produce thermonuclear explosions, produces supernova light-curves that
are broader than the observed type Ia sample.  In addition, we 
discuss how the shock breakout and spectral features of these double 
degenerate progenitors will differ from the canonical bare 
Chandrasekhar-massed explosion models.  We conclude with a 
discussion of how one might reconcile these differences with 
current observations.

\end{abstract}


\section{Introduction}

Type Ia supernovae are among the strongest explosions observed by
astronomers.  These thermonuclear bombs dominate the production of
many iron-peak elements in the universe.  Their near-uniform and
easily calibrated light-curves have allowed astronomers to use them
as probes into the universe and the nature of cosmology.  However,
despite their extreme importance, we still do not understand the
details of their explosion mechanism nor do we understand the exact
nature of their progenitor.  Although it has been argued that white
dwarf systems with masses below the Chandrasekhar limit may produce
thermonuclear explosions and, in WD-WD collisions, this is generally 
true \citep{Rask09, Ross09}, it is generally believed that most type
Ia supernovae are produced as an accreting WD is pushed above the
Chandrasekhar mass limit (see Livio 2001 for a review).  In this
paper, we study the progenitors of these Chandrasekhar-massed
explosions.

The exact nature of the binary progenitor that leads to the WD
accretion is also a matter of debate.  The wide range of type Ia
progenitors have been divided into two clases (see Livio et al.  2000
for a review): double degenerate mergers where the white dwarf
accretes material during the merger with a white dwarf companion, and
the single degenerate scenario where the white dwarf accretes material
from a normal star (typically giant star) companion.  Although
simulations of double degenerate mergers suggest that some of these
mergers produce type Ia supernovae \citep{Yoon07,Pak10}, it is
believed that most of these mergers will form accretion-induced
collapse implosions instead of type Ia supernovae
\citep{NK91,NUK01,Liv01}.  Based on stellar accretion models, we would
assume that the primary progenitor of type Ia supernovae is the single
degenerate accretion scenario.

However, population synthesis calculations currently argue against the
single-degenerate scenario.  To date, population synthesis
calculations predict roughly an order of magnitude more double than
single degenerate scenarios (see Livio 2001 for a review).  The double
degenerate merger rate is consistent with the observed supernova rate 
(hence, the single-degenerate rate is an order of magnitude too low 
to explain observations).

The rate is not the only observational constraint we have on the
progenitor of type Ia supernovae.  The spectra from supernova
outbursts provide clues as well.  H\"oflich (2005) has used the
light-curve and spectral observations of supernovae to constrain the
progenitor, the environment and the details of the explosion.  Among
these, one can constrain the level of asymmetry and the mass of the
white dwarf.  Both sub- and super-Chandrasekhar white dwarfs produce
features (e.g. slow rise/decline in the super-Chandrasekhar white
dwarfs) that differ from standard type Ia supernovae, arguing that
most Ia explosions are produced in the explosion of a
Chanrasekhar-massed white dwarf\citep{Maz07}.  In this paper, we
combine population synthesis studies of binary mergers, hydrodynamical
models of mergers and radiation-hydrodynamics calculations to study
the observational implications for the double degenerate Type Ia
progenitor scenario.  This combined theoretical effort allows us to
make definitive observational predictions for the double degenerate
scenario.  Comparing these predictions with observations can place
limits on the fraction of Type Ia supernovae produced by double
degenerate progenitors.

The codes used in this paper have been detailed in other papers.  
In Chapter 2, we briefly describe the codes used and outline how 
we couple the codes together.  This coupling of codes allows us 
to make stronger predictions than have been made in the past.  
Most important is the coupling of stellar merger models, 
guided by population synthesis models, to our light-curve 
calculations.  We describe the fate of these mergers and 
how they couple to light-curve calculations in Chapter 3.  
The environment created by these mergers drastically affects 
the emission from these explosions.  In chapter 4 we study 
the effects of this environment on the initial shock emergence.  
Chapter 5 discusses the light-curves and spectra around peak 
for these explosions.  We conclude with comparisons 
to supernova observations and their subsequent constrants 
on the double degenerate scenario.

\section{Code Description}
\label{sec:code}

Our populations synthesis studies are from the simulation results of
Ruiter et al. (2009) which used the StarTrack population synthesis
code.  A detailed description of the algorithm and the input physics
is given in Belczynski et al. (2002, 2008).  Ruiter et al. (2009)
found that, assuming all CO$+$CO mergers above a Chandrasekhar mass
produced Ia supernovae, the double-degenerate scenario dominated the
total, time-integrated Ia rate (although, depending upon the model,
the rate of single degenerate mergers can be comparable after about
2\,Gyr after a burst of star formation).  Here they assumed ONe white
dwarfs above a Chandrasekhar mass collapsed to form neutron stars and
there were no CO-He mergers with masses above the Chandrasekhar limit.
In this paper, we use the distributions of the total system mass from
these simulations to guide our type Ia progenitors.

Based on the masses in our population synthesis, we model the merger
of two CO cores using the SNSPH code~\citep{Fry06}.  This code follows
the evolution of the merger from the onset of Roche-lobe overflow
through the disruption of the companion white dwarf.  These
simulations provide us with a realistic description of the
distribution of matter surrounding the white dwarf after the dynamical
phase of the merger.  Analytic estimates of the cooling (including
convection) in this post-merger structure allows us to place limits on
the density structure at the time of explosion.  We use this range of
structures to calculate the range of light-curves and spectra we
expect from these systems.

\begin{figure}
\plotone{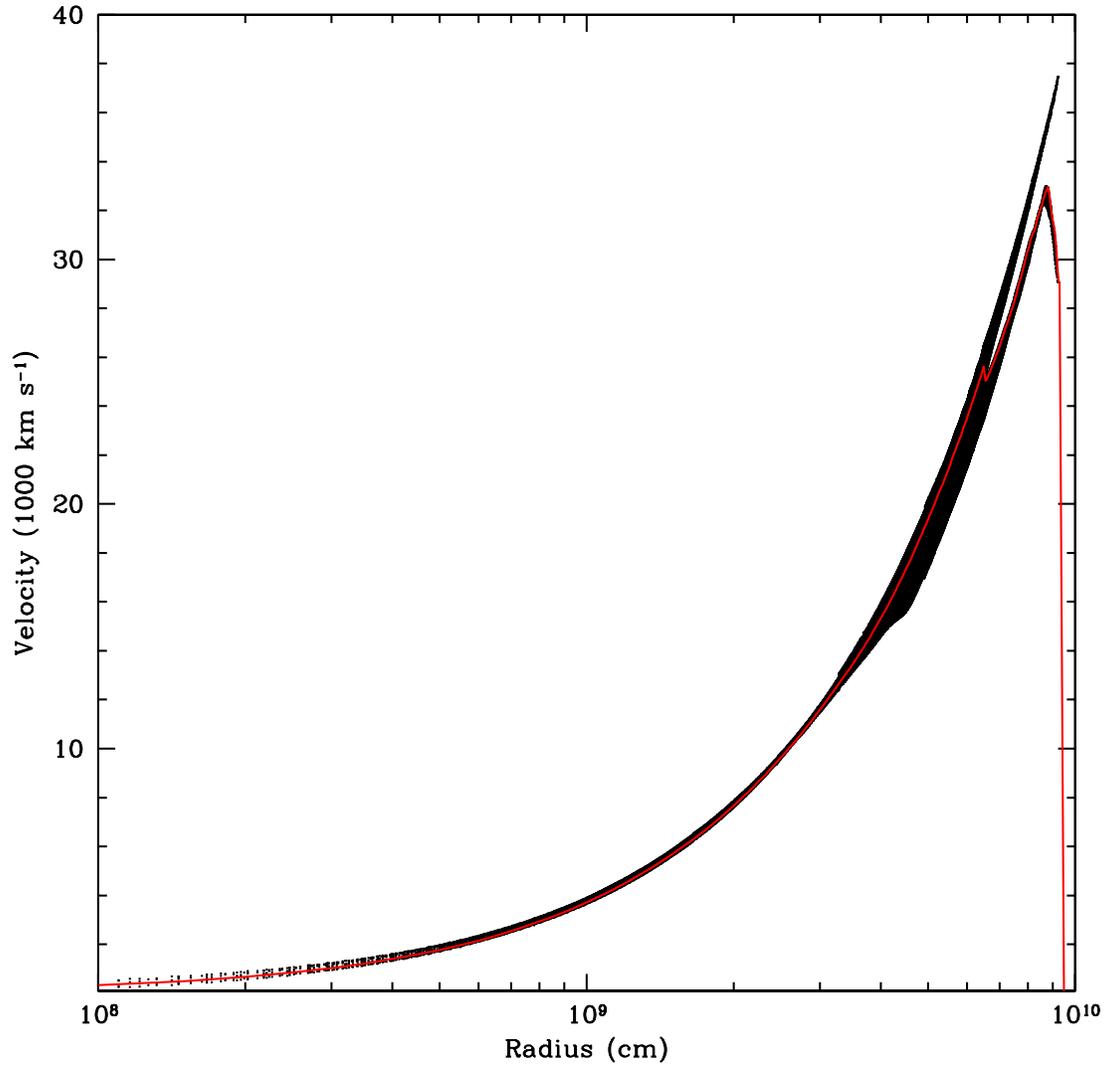}
\vspace{-1.8in}
\caption{Velocity profile of our type Ia supernova explosion, 
based on simulations by Meakin et al. (2009).  The black points depict the 
velocity versus radius from this simulation.  The line shows our 
1-dimensional fit to these data.  Our 1-dimensional fit can not 
match the full structure of this explosion, but it does match 
the basic structure and conserves the total energy.}
\label{fig:explosion}
\end{figure}

\begin{figure}
\plotone{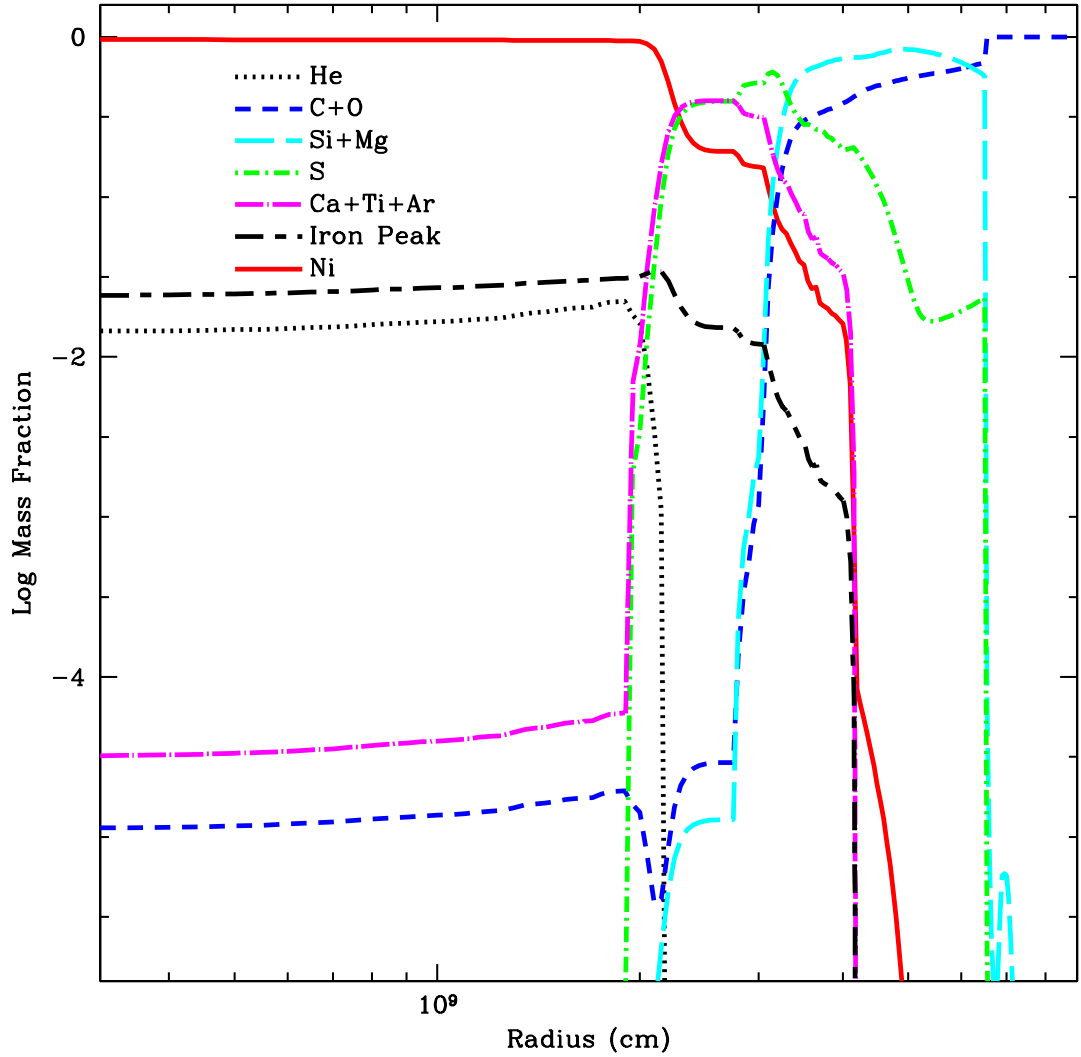}
\vspace{-1.8in}
\caption{1-dimensional abundance profile of our type Ia supernova
explosion (see Fig.~\ref{fig:explosion} for details).  We have combined
the yields to focus on the major trends in this distribution profile.
For our opacities we use 14 representative elements.}
\label{fig:abun}
\end{figure}

Our supernova is produced in a gravitational confined detonation model
\citep{Ple04} from the Chicago FlASH team \citep{Mea09}.  The
explosion simulation uses the FLASH code \citep{Fryx00} and is modeled
in 3-dimensions, which we have then mapped into 1-dimension.  A
comparison of our 1-dimensional explosion velocity profile on top of
the 3-dimensional profile is shown in figure~\ref{fig:explosion}.
Although we are modeling the general radial profile of this explosion
with our mapping, this explosion is clearly multi-dimensional and any
comparison of this explosion model to data will require
multi-dimensional light-curves.  However, this explosion provides a
modern explosion with which to do our comparisons.  The corresponding
yields for our 1-dimensional mapping are shown in
figure~\ref{fig:abun}.  We reduced the yields to 14 representative
elements for our opacities: H, He, C, O, Ne, Mg, Si, S, Ar, Ca, Ti,
Cr, Fe, and $^{56}$Ni.  The total $^{56}$Ni yield in this explosion is
0.7M$_\odot$.  For most of the models, the explosion energy is 
roughly $1.6 \times 10^{51}$\, erg, but we included a series of weak 
explosions as well $0.4 \times 10^{51}$\, erg.

The results from the above codes are all brought together in our
calculations of the emission using the LANL supernova light-curve
code~\cite{Fry09}.  These calculations couple a radiation-hydrodynamic
simulation using the RAGE \citep{Fry07,Git08} code with post-process
spectra.  Both codes use the LANL OPLIB database \citep{Mag95} for
opacities.  Although the RAGE code is capable of running with a large
set of photon energy groups, the current calculations use a single
group Rosseland-mean averaged opacity using the OPLIB database.  For
the post-process calculations, we use the full frequency-dependent
opacity information (14,900 groups) from this database.  The
1-dimensional structure is mapped into a multi-dimensional profile
(each zone is divided into 80 angular bins).  By choosing a
line-of-sight and calculating the emission along the entire exploding
star, we model the entire ejecta including limb effects.  This
approach also allows us to calculate the full Doppler effect of the
expanding material, including red- and blue-shifted opacities.  For
more details, see \cite{Fry09}.  With these calculations, we can
compare our different merger environments to predict observational
trends between these models.

\section{Stellar Merger and Subsequent Accretion}
\label{sec:merger}

In this paper, we use the results from the population synthesis models
from Ruiter et al. (2009) to guide our initial progenitor mass
distribution.  Figure~\ref{fig:popsynth} shows the results of three
simulations using different parameters for the common envelope
evolution: model 1: $\alpha$ prescription
\footnote{The common envelope follows a prescription: $\alpha_{\rm CE}
  \left( \frac{G M_{\rm don,fin} M_{\rm primary}}{2 A_{\rm fin}} - \frac{G
    M_{\rm don,int} M_{\rm primary}}{2 A_{\rm int}} \right) = \frac{G
    M_{\rm don,int} M_{\rm don,env}}{\lambda R_{\rm don,lob}}$ where $G$ is
  the gravitational constant, $M_{\rm primary}$, $M_{\rm don,int}$ and
  $M_{\rm don,fin}$ are the mass of the primary, the initial mass of
  the donor and final mass of the donor stars respectively, $A_{\rm
    int}$ and $A_{\rm fin}$ are the initial and final orbital
  separations, $R_{\rm don,lob}$ is the donor Roche lobe radius and
  $\lambda$ is a measure of the central concentration of the donor
  (Webbink 1984).}  where $\alpha \times
\lambda = 1$, model 2: $\alpha$ prescription where $\alpha \times
\lambda = 0.5$, model 3: $\gamma$ prescription where $\gamma$=1.5.
The $\gamma$ prescription of model 3 is similar to that of Nelemans \&
Tout (2005)\footnote{The common envelope is treated as a evolution of
  the angular momentum as mass is lost from the system: $\Delta J/J =
  \gamma \Delta M/M$ where $\Delta J$ and $\Delta M$ are the change in
  the angular momentum and mass respectively, $J$ and $M$ being the
  initial angular momentum and mass.  $\gamma$ is the efficiency at
  which angular momentum is lost with mass \citep{Kool90,Nel00,Nel05}.};
the difference being that we assume this $\gamma$-formalism for common
envelope evolution every time a common envelope event is encountered.
Nelemans \& Tout (2005) use both a $\gamma$ and $\alpha$ depending on
the nature of the common envelope (the $\alpha$-formalism is never
used in our population synthesis model 3).
  
There are a number of different formation channels producing CO$+$CO
mergers.  Changing the common envelope efficiency and/or prescription
changes both the rate and the relative importance of these channels.
Overall, our model 1 produces the most massive CO$+$CO mergers.  Model
2, with its lower common-envelope removal efficiency, produces smaller 
orbits after the first common envelope phase.  Many of these systems
merge before the formation of the second white dwarf (see Ruiter et
al. 2009 for details).  These two models are standard models from
Ruiter et al. (2009) and we will focus on their results in this paper.
Although the two simulations predict different rates and merger times,
they have several similarities.  Both mass distributions are double
peaked, one $\sim 1.05$M$_\odot$, the other $\sim 2$M$_\odot$.  If we
restrict ourselves to only those systems with total masses above the
Chandrasekhar limit, over a third of the systems lie in the peak at
$\sim 2$M$_\odot$.

Our population synthesis caluclation using the $\gamma$ prescription
(model 3), however, produces even fewer massive CO$+$CO mergers than
our $\alpha$ prescription models.  This is mainly due to the fact that
the dominant formation channel which contributes to CO$+$CO white
dwarf mergers at M$_{\rm tot}>\sim 1.8$M$_{\odot}$ in the first two
models is nearly absent in model 3 (producing no ``secondary spike''
of events $\sim 2$M$_{\odot}$).  We will discuss this further in
chapter~\ref{sec:summary}.

\begin{figure}
\plotone{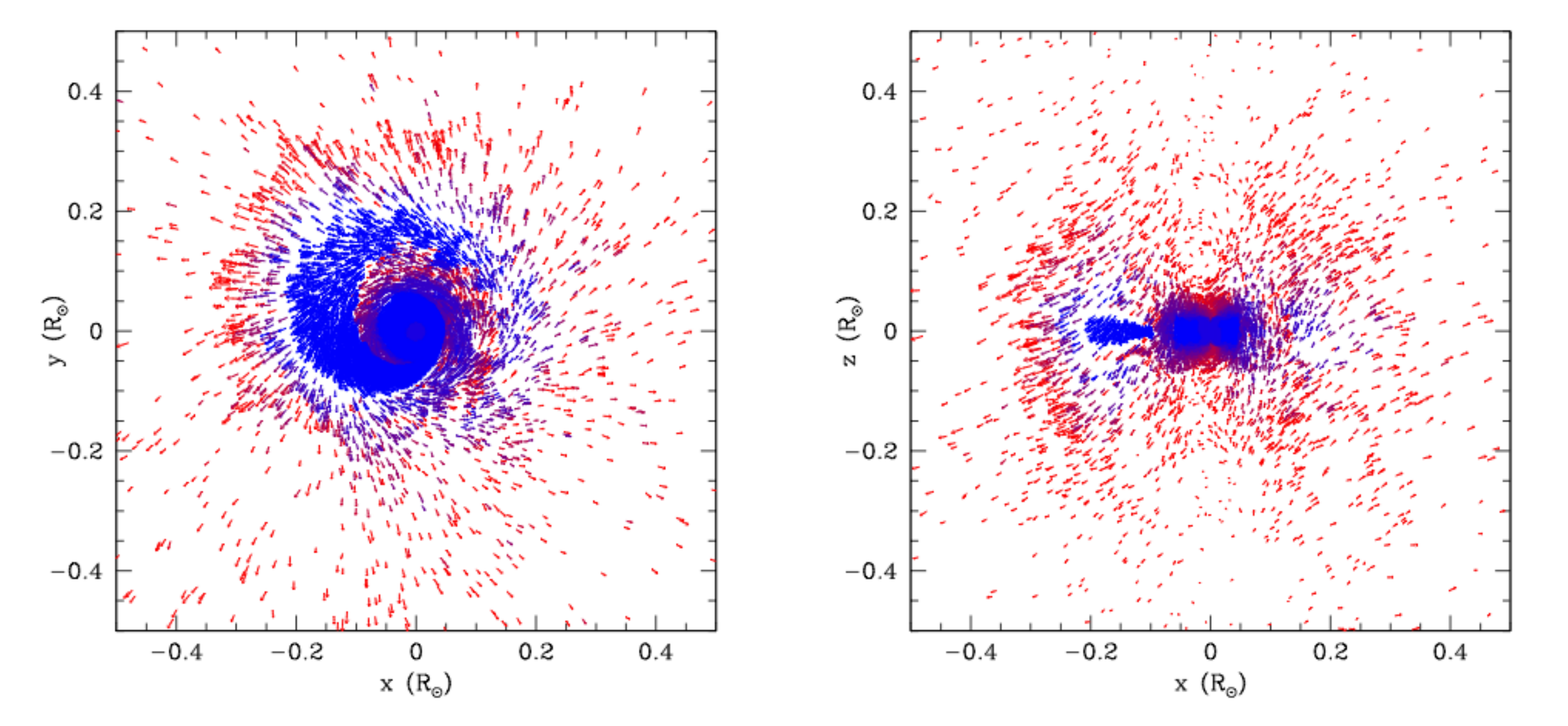}
\caption{A slice in the x-y (orbital) and x-z (out of orbital) plane
  of our 3-dimensional simulations of white dwarf mergers (a
  simulation using the SNSPH code).  The coloring denotes density
  (blue is high, red is low) and the vectors denote velocity magnitude
  and direction.  Although far from symmetric, we will angle average
  this density profile for our 1-dimensional explosions.  Clearly,
  multi-dimensional models are required to produce accurate density
  profiles and, presumably, spectra and light-curves for the supernova
  in these systems.}
\label{fig:mergexy}
\end{figure}

\begin{figure}
\plotone{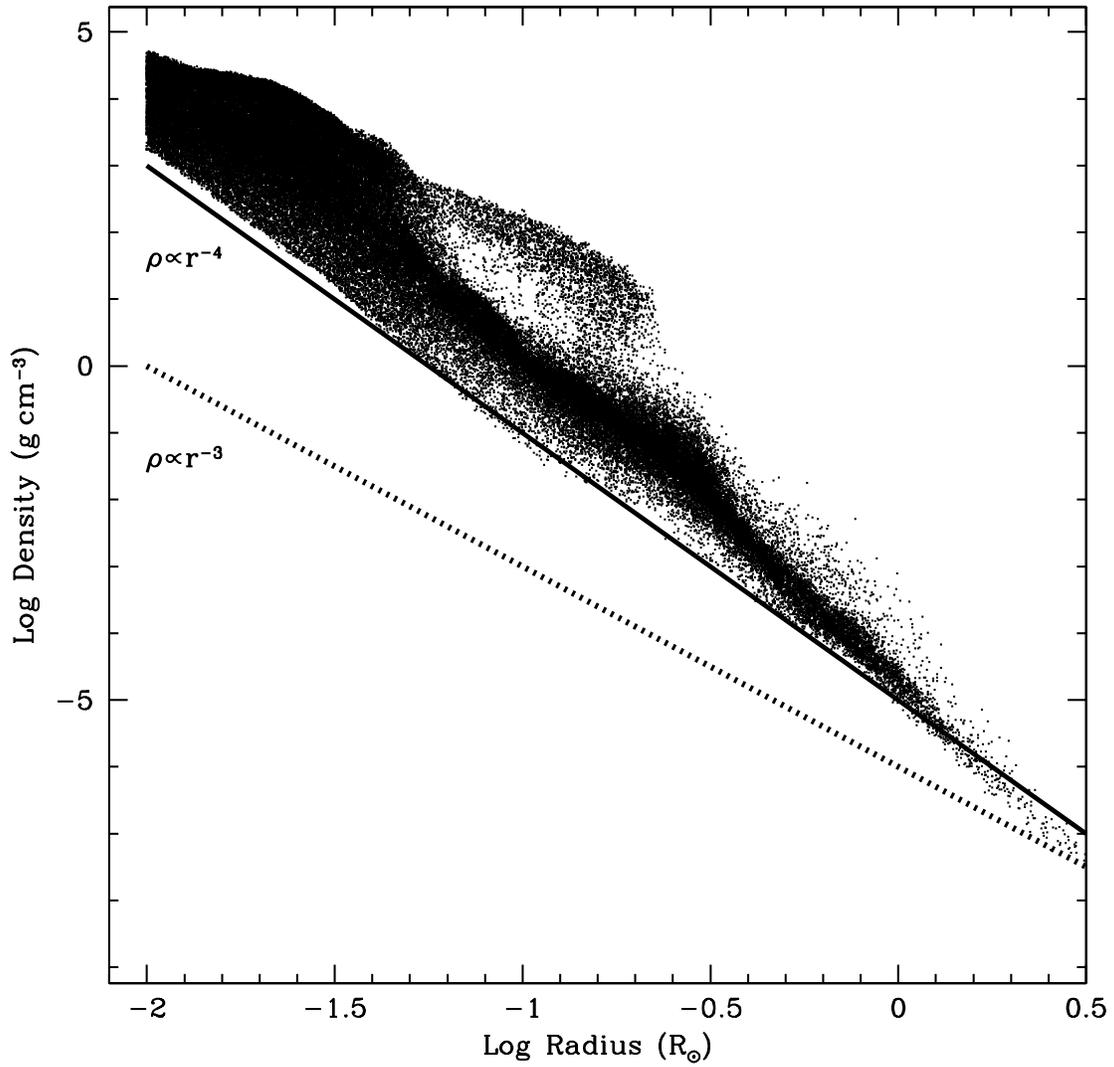}
\vspace{-2.0in}
\caption{The density profile of the outer envelope 
of our star.  Although there is considerable structure 
in this profile, it is fit reasonably well with an 
r$^{-4}$ density structure.  We will assume this 
power-law for most of our simulations.}
\label{fig:mergedens}
\end{figure}

With the peak in total system masses around $2$M$_\odot$, we modeled
the merger of a 0.9\,$M_\odot$ WD with a 1.2\,$M_\odot$ WD.  We follow
this merger until the less-massive white dwarf is completely
disrupted.  This leaves behind a compact core with an extended merger
envelope (Fig.~\ref{fig:mergexy}).  As with our explosion model, the
structure is highly aspherical.  Nonetheless, for this study, we map
this structure into a 1-dimensional profile
(Fig.~\ref{fig:mergedens}).  As the core accretes above the
Chandrasekhar limit, it will collapse (and possibly produce a type Ia
explosion).  Above this white dwarf is the rest of the debris from the
merger.  The density profile of this debris has a good deal of
structure, but in 1-dimension, it is reasonably well-fit by: $\rho =
\rho_0 (r/r_0)^{-4}$ where $r$ is the radius from the center and
$\rho_0$ is the density at radius $r_0$.

For the system modeled in our merger, the inner 1.4\,M$_\odot$ will
accrete onto the white dwarf fairly quickly.  By using the conditions
at the end of our merger calculations, we estimate the cooling time
(set to $E_{rad}/L_{rad}$ where $E_{rad}$ is the energy in this inner
atmosphere, dominated by radiation, and $L_{rad}$ is the radiative
flux out of this inner material) to be $ < 10,000$\,s.  As this
material cools, it accretes onto the white dwarf, ultimately pushing
it above the Chandrasekhar limit.  If this produces a thermonuclear
explosion, it will drive a shock through the remaining 0.7\,M$_\odot$.
Due to the short cooling time of the inner material, the density
profile of this outer 0.7\,M$_\odot$ will be very similar to its
initial state (its cooling time is many orders of magnitude longer).
For our light-curve calculations, we use this $r^{-4}$ profile.  With
less massive mergers, the characteristics of the density profile might
change slightly, but this $r^{-4}$ profile should not change.  We use
this profile for the bulk of our simulations.  If the time-scale is
much longer and the atmosphere reaches an equilibrium, entropy-driven
instabilities will produce a constant entropy
atmosphere~\citep{Hou92,Col93}.  For these compact object systems, the
atmosphere is reasonably well described by an $r^{-3}$ profile and this
represents the shallow extreme of the possible density profiles.  To
test the dependence of our results on this choice, we model one
calculation with an $r^{-3}$ profile.

Assuming this power law density distribution, we can then set the
initial conditions of our models as a function of total envelope mass.
For our light-curves, we use 3 different envelope masses: 0.1, 0.35,
and 0.7\,M$_\odot$.  This corresponds to total binary masses of 1.5,
1.75, and 2.1\,M$_\odot$.  The companion star could be a He white
dwarf or a CO white dwarf.  We set the composition to these two
separate abundances (helium vs. carbon and oxygen)\footnote{These
  helium masses are not included in our population synthesis models
  and the higher masses are not physical.  But these models give us an
  idea of how composition can affect the spectra and light-curves.}.
The entire suite of models is summarized in
Table~\ref{tab:sims}\footnote{This table also includes many of the
  observational results discussed in chapter~\ref{sec:spectra}.}.

\section{Shock Emergence}
\label{sec:emergence}

Our radiation-hydrodynamics simulations are ideally suited to
calculating the initial emergence and early evolution (through peak)
of these buried type Ia supernovae. The radiation-hydrodynamics
simulations allow us to include any additional shock heating as the
exploding white dwarf interacts with its surroundings.  Simulations
that calculate radiative transfer on a homologous outflow scheme will 
not include this shock heating.  In addition, at shock breakout, the 
material temperature is not set entirely by the radiation.  This 
disequilibrium betwen radiation and matter can not be modeled by 
pure radiation calculations.

We use the density/temperature profiles of these calculations to
provide input for our detailed ($>10,000$ energy group) calculations of the
spectra.  With these spectra as a function of time, we can calculate
the light-curves in any energy band (see Fryer et al. 2009 for
details).

As the type Ia supernova expands, the density at the head of the shock
decreases.  The first visible emission of the supernova explosion
occurs when this density becomes so low that radiation is no longer
trapped in the flow.  H\"oflich \& Schaefer (2009) recently studied
this emergence of the supernova shock.  For a normal type Ia
supernovae, they argued that this occurs when the shock is roughly at
9000\,km.  At shock emergence, the type Ia emits a burst (up to
$10^{50}$\,erg\,s$^{-1}$) of X-ray and gamma-ray radiation.  Piro et
al. (2010) argue that the emission peaks closer to
$10^{44}$\,erg\,s$^{-1}$.  In both cases, the emission has dropped 
considerably by 1\,s.

This shock emergence is similar to shock breakout of type Ib/c and II
supernovae in that the initial emission is driven by the thermal
energy in the shock.  The temperature in the shock tends to be high,
resulting in an initial burst of high-energy photons.  The primary
difference is that the progenitor stars of type II supernovae are more
extended and the radiation remains trapped until the shock breaks out
of the star ($10^{7}-10^{9}$\,km).  Type Ib/c stars are more compact,
but their strong winds are sufficiently dense to trap the radiation
until the shock is further out as well.  At these higher radii, the
thermal temperature is much lower, and the breakout emission tends to
be at much shorter wavelengths (UV and X-ray bands).  The
corresponding emission timescale is longer (Frey et al. 2010).

\begin{figure}
\plotone{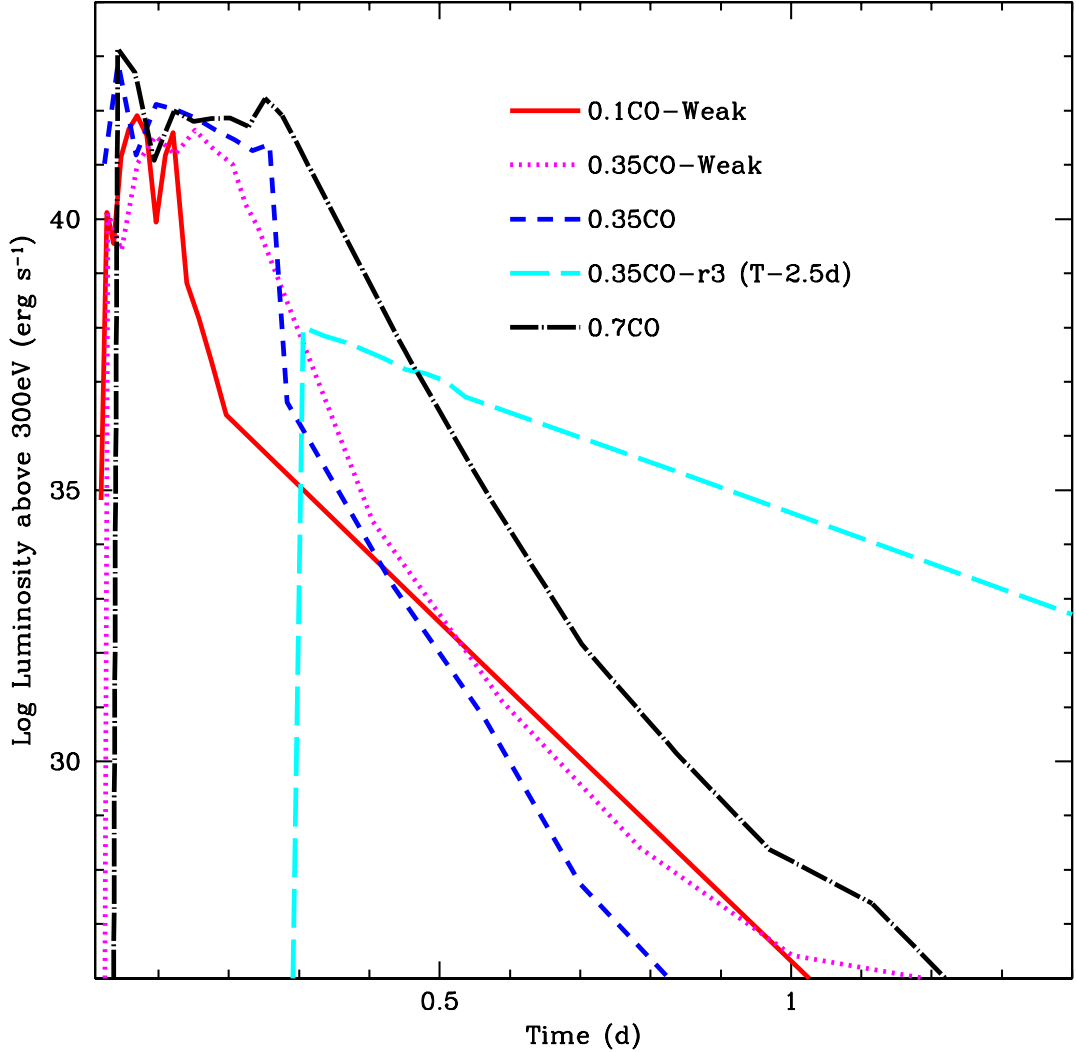}
\vspace{-1.8in}
\caption{The X-ray luminosities as a function of time from shock
  break-out for 5 of our CO surrounding atmospheres: 0.1CO-Weak
  (solid), 0.35CO-Weak (dotted), 0.35CO (dashed), 0.35CO-r3
  (long-dashed), 0.7CO (dot-dashed).  Most of our models peak around
  $10^{43}$\,erg\,s$^{-1}$ and the emission lasts for 0.1-0.3 days.
  The exception is the atmosphere where we used the $r^{-3}$ density
  profile.  In this model, the radiation remains trapped longer and
  shock break-out occurs at higher radii when the shock is cooler.
  This produces a lower X-ray flux with a longer duration.  All of
  these models are very different from the shock emergence seen in
  normal Ia supernovae, which have no surrounding
  atmospheres~\citep{Hof09,Pir10}.}
\label{fig:xray}
\end{figure}

The type Ia SNe from WD mergers studied in this paper are very similar
to the shock breakout in Type II supernovae.  The merger debris acts
similarly to a Ib/c wind, trapping the radiation until the shock is
well beyond the white dwarf itself ($\sim 10^{13}$\,cm for our steep,
$r^{-4}$ density profiles and $>10^{14}$\,cm for the shallower
$r^{-3}$ profiles).  For explosions on a bare white dwarf, the
radiation becomes untrapped when the shock breaks out of the white
dwarf, roughly $10^9$\,cm \citep{Hof09}.  This leads to a longer
emission timescale and a smaller peak emission in the X-ray than that
predicted for normal (``bare'') Type Ia
supernovae~\citep{Hof09,Pir10}.  Figure~\ref{fig:xray} shows the total
emission above 300eV for 5 of our models.  In most cases, the peak in
the X-ray flux arises within half a day of the explosion.  The peak
emission is roughly 10$^{42}$\,erg\,s$^{-1}$ and this peak lasts
10,000-30,000\,s (0.11-0.35\,d).  As the surrounding material
increases, so does the duration of the X-ray emission.  But the
structure of the density profile plays a large role in determining the
peak flux and duration.  Note that the model with 0.35\,$M_\odot$
surrounding material with the $r^{-3}$ dependendence on the density
has a much later peak (2.8\,d after explosion) and a peak flux of only
$10^{38}$\,erg\,s$^{-1}$.

\begin{figure}
\plotone{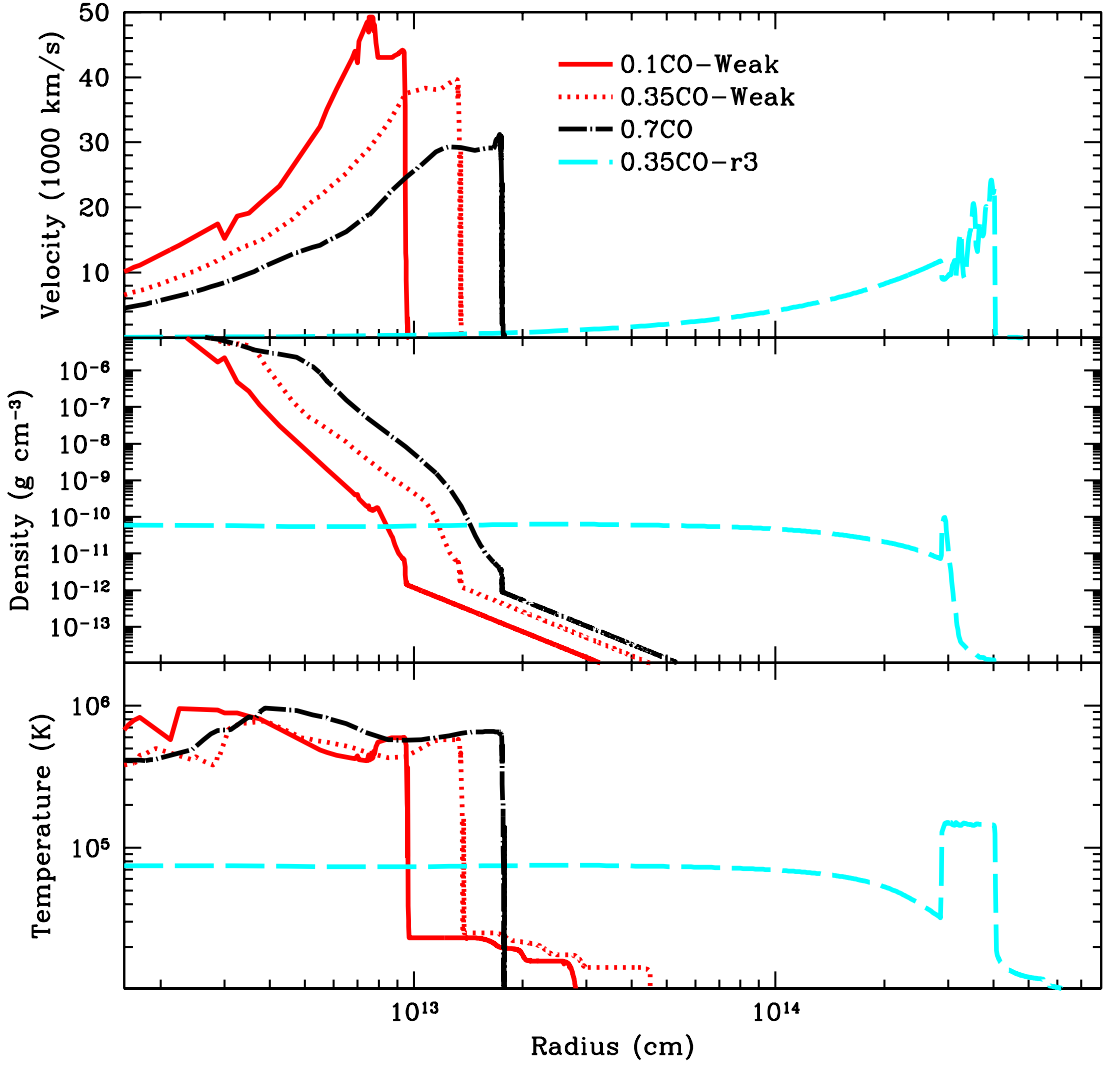}
\vspace{-1.8in}
\caption{The velocity (top), density (middle), and temperature
  (bottom) profiles at shock break-out of 4 of our shock
  models: 0.1CO-Weak (solid), 0.35CO-Weak (dotted), 0.35CO-r3
  (long-dashed), 0.7CO (dot-dashed).  Shock breakout occurs later for
  more massive envelopes.  The shallower density gradient for model
  0.35CO-r3 traps the radiation longer.  Shock break-out for this
  model occurs after the shock has traveled 15-30 times further out.
  This explains the later X-ray emergence and shallower decay in the
  X-ray luminosity.  Its temperature is correspondingly lower at
  shock-breakout, leading to a lower X-ray luminosity.}
\label{fig:xraystruc}
\end{figure}

These trends are easy to understand based on the mass of the
surrounding atmosphere.  Figure~\ref{fig:xraystruc} shows the
velocity, density and temperature structures of 4 of our models at
shock breakout.  We expect the breakout to occur when the shock hits
the photosphere (or roughly when the mean free path is on par with the
width of the shock).  The width ($\Delta r$) of the shock grows with
time, roughly staying a constant fraction of the shock size or
position: $\Delta r \propto r_{\rm shock breakout}$.  The mean 
free path is given by $(\sigma \rho)^{-1}$ where $\sigma$ is the 
opacity in $cm^2 \, g^{-1}$.  Setting the mean 
free path equal to the shock width and using our density profile 
($\rho = kr_{\rm shock breakout}^{-4}$ where $k$ depends linearly 
on the envelope mass) , we find:
\begin{equation}
r_{\rm shock breakout} \propto r_{\rm shock breakout}^4/(k \sigma).
\end{equation}
Solving for $r_{\rm shock breakout}$, we obtain:
\begin{equation}
r_{\rm shock breakout} \propto (k \sigma)^{1/3}.
\end{equation}
The ratio of the shock breakout radii ($r_{\rm shock breakout,
  M_1}/r_{\rm shock breakout, M_2}$) for two different envelope masses
($M1,M2$) is simply:
\begin{equation}
r_{\rm shock breakout, M_1}/r_{\rm shock breakout, M_2} = (k_{\rm
  M1}/k_{\rm M2})^{1/3} = (M1/M2)^{1/3}.
\end{equation}
We expect, then, that the
shock in the 0.7\,M$_\odot$ model to experience shock breakout roughly
1.9 times further out than the 0.1\,M$_\odot$ model.  The duration of
the breakout should be roughly proportional to $r/v_r$ where $v_r$ is
the shock velocity.  Hence, we also expect longer durations for our
more massive atmospheres (From figure~\ref{fig:xraystruc}, we see that
both the radius is bigger and the radial velocity is smaller).

The primary outlier in this picture is the shallower density profile
model.  But the same physics explanations apply.  The shallower
density means that shock breakout occurs at much higher radii.  Not
only is the breakout delayed (and the duration of the breakout is
longer), but at these large radii, the shock has cooled, leading to 
a much weaker signal.  To accurately calculate the breakout emission, 
we will need to have accurate estimates of the density profile of the 
surrounding material.

Can we use shock breakout emission to constrain the progenitor?  If
the breakout emission is extended for more than 1000\,s, we are
assured that non-neglibile material lies on top of the exploding white
dwarf.  Technically, the duration scales with the amount of mass on top.  
But as we have found from altering our density profile, the profile is 
as important, if not more important, than the total amount of mass.  
Until we have accurate models of the merger process and can model 
the explosions through these profiles in multi-dimensions, it will 
be difficult to constrain the exact mass of the system from shock 
breakout alone.

\section{Spectra and Light-Curves Near Peak}
\label{sec:spectra}

We are not limited solely to the shock breakout light-curve.  The 
surrounding material will also affect the light-curve and spectra 
near peak.  Especially for these enshrouded type Ia, shock heating 
remains an important feature of the light-curve through peak and our 
radiation-hydrodynamics calculations are ideally suited for calculating 
spectra and light-curves of these supernovae.

\begin{figure}
\epsscale{0.75}
\plotone{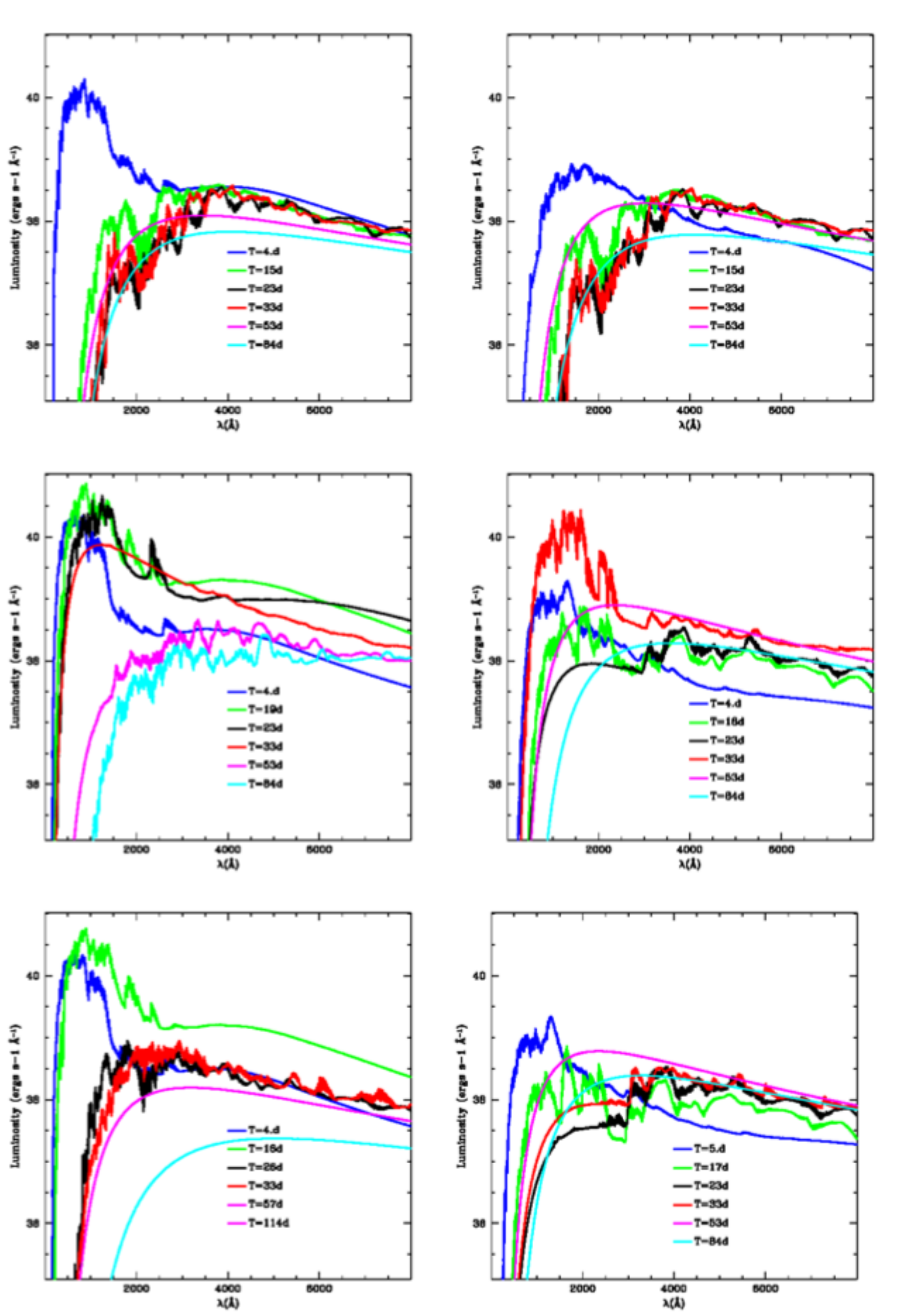}
\epsscale{1.0}
\caption{Spectra (luminosity versus wavelength) from our 6 primary
explosion models at a series of time positions.  The left column
corresponds to helium atmospheres, the right column corresponds to CO
atmospheres.  The envelope masses range from 0.1 to 0.35 to
0.7\,M$_\odot$ for the top, middle, and bottom panels respectively.
At early times, there is a peak at low wavelengths (high energies)
denoting the initial emergence of the radiation.  With time, this peak
moves to lower energies (longer wavelengths).  Ultimately, the
material becomes so diffuse that it is ionized by the radiation and
many of the red line features disappear.}
\label{fig:fullspec}
\end{figure}

\begin{figure}
\epsscale{0.75}
\plotone{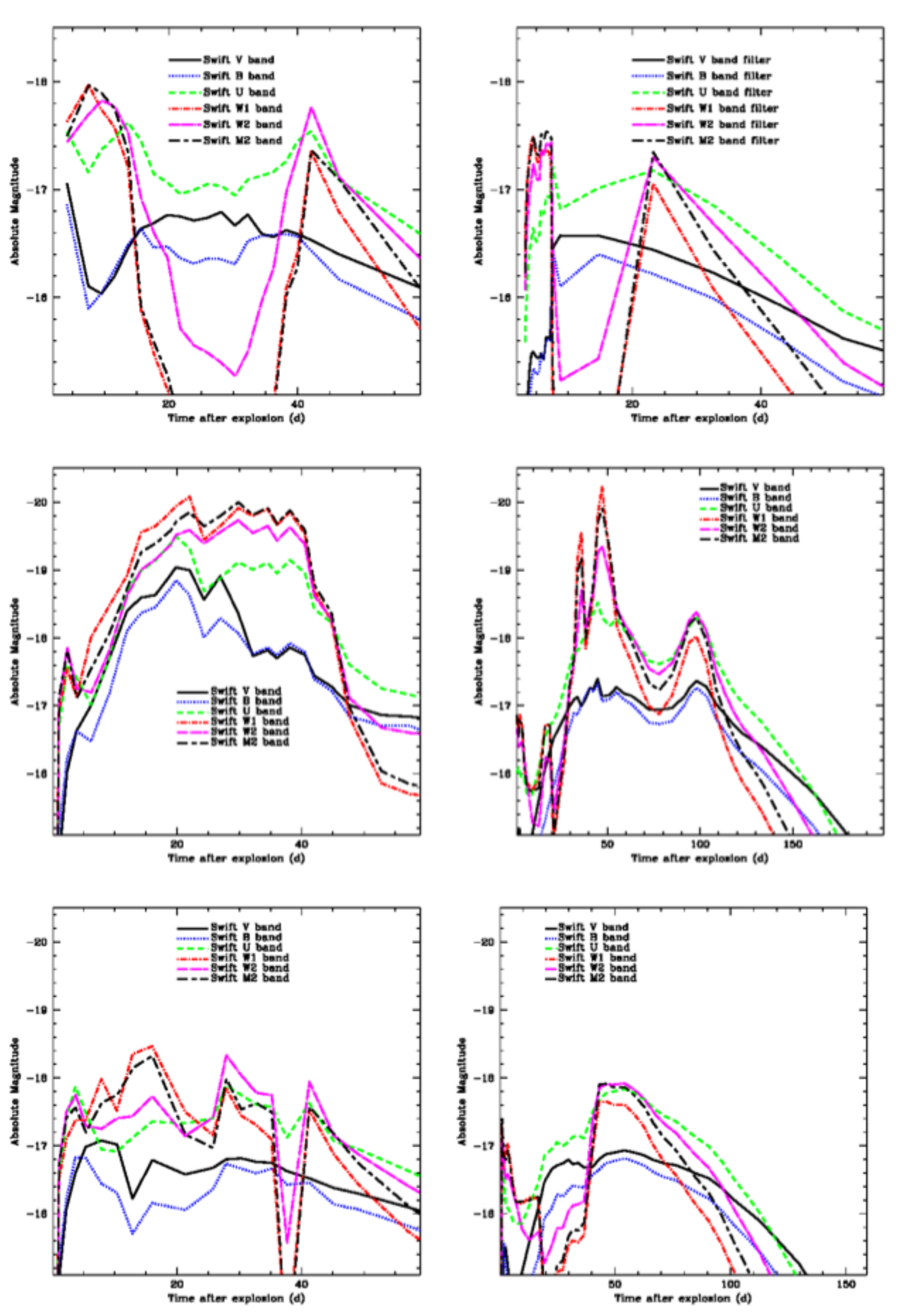}
\epsscale{1.0}
\caption{Light-curves (magnitude versus time) of our 6 primary models
for a range of filter bands (See Fig.~\ref{fig:fullspec} for details).
Here we are using the filters used in the Swift satellite.  Shock
heating in the envelope produces extended UV emission and may also
broaden the emission in the V-band.  As the envelope mass increases,
the emission is first increased (as shock heating becomes more
important), and then decreases as photon trapping limits and delays
the emission.}
\label{fig:fulllc}
\end{figure}

Our basic suite of light-curve models consists of 3 different envelope
masses: 0.1, 0.35, and 0.7\,M$_\odot$, corresponding to total binary
masses of 1.5, 1.75, and 2.1\,M$_\odot$.  For each envelope mass, we
model two simulations, one assuming the envelope is composed of carbon
and oxygen, the other assuming it is all helium (see
Table~\ref{tab:sims} for details).  In this way, we study not only the
double degenerate progenitor consisting of two CO cores, but also the
dependence on the composition of the debris.
Figure~\ref{fig:fullspec} shows the spectrum over a broad wavelength
and flux range for each of our 6 models.  The corresponding 
light-curves for a range of wavelength filters are shown in 
Figure~\ref{fig:fulllc}.

Let us review the trends in the light-curves first.  One pervasive
feature of a surrounding environment is that it can delay initial
photon emergence.  The surroundings trap the photons longer, delaying
the UV burst.  In addition, shock heating as the explosion hits this
surrounding environment produces high temperatures and longer (or
second) UV outbursts.  This means that the UV outburst may still be
strong at peak V-band emission.  A large flux shorter than 1500\,$\AA$
at peak V-band is a strong indicator of a large surrounding
environment and, hence, a double-degenerate progenitor.  If we further
increase the mass of the surrounding medium, the radiation is trapped
longer.  This will produce a broader light-curve.  But shock heating 
plays a larger role in driving the supernova, and this extra energy 
source makes it more difficult to analyze the light-curve.

Many of these trends can be seen in figure~\ref{fig:fulllc}.  With a
0.1\,M$_\odot$ surrounding environment, the light-curves of the helium
and carbon/oxygen mergers are very similar.  The V-band peaks between
15-30\,d with an absolute magnitude dimmer than -17.  The UV emission
is initially bright at first emergence from the expanding ejecta and
then peak again due to shock heating (20-40\,d).  The peak UV emission
predicts absolute magnitudes near -17 in all the Swift bands.

As the environment mass increases, so does this peak emission.  
With a 0.35\,M$_\odot$ surrounding environment, the UV emission 
peaks at absolute magnitudes of -20.  Shock heating is dominating 
this UV emission.  For a helium envelope, where the opacity is 
lower because the helium in the envelope is quickly ionized, the 
UV emission is broadened and peaks at -20.  The V-band also 
peaks higher (-19), but decays in 10\,d by over a magnitude.  
For C/O envelopes with their higher opacities, the UV peaks 
high (-20) at later times than its helium counterpart.  
In addition, the V-band is strongly broadened.  Even after 
60\,d, it is within 0.2 magnitudes of its peak.

At still higher envelope masses (0.7\,M$_\odot$), photons are trapped
longer.  When photons do escape, the temperature of the ejecta is
lower, leading to lower UV emission than in our 0.35\,M$_\odot$ case.
Even for helium surroundings, the V-band emission begins to produce a
broader peak.  But this broadening is exacerbated in the high-opacity
C/O surroundings.  For massive C/O envelopes, the V-band is still
rising after 60\,d.

\begin{figure}
\plotone{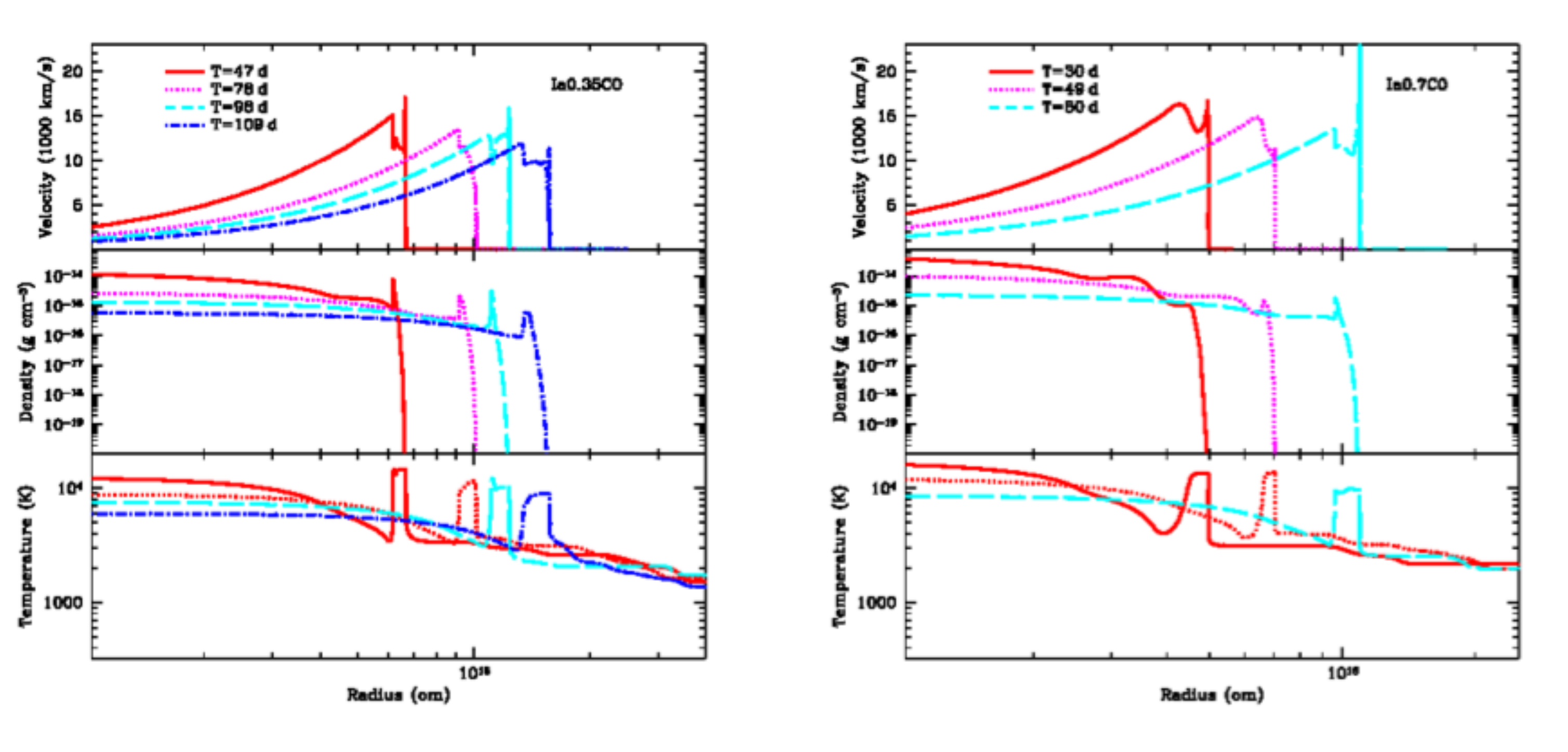}
\caption{The velocity (top), density (middle), and temperature
  (bottom) profiles near shock break-out at a variety 
of times for our more massive CO envelope models.  Throughout the 
peak emission for both models, the shock remains at the boundary 
between trapped radiation and free-streaming radiation.}
\label{fig:struct}
\end{figure}

The behavior of these light-curves is very different from those of
normal Type Ia supernovae.  These differences, in part, can be
understood by the differences in the structure of the star throughout
peak.  Figure~\ref{fig:struct} shows the velocity, density and
temperature profiles of our standard 0.35\,$M_\odot$ and
0.7\,$M_\odot$ envelopes at a range of times during the peak breakout
emission.  Note that the shock remains at the boundary between trapped
flow and free-streaming throughout peak.  The front of the shock is
pushing through low-density media and emitting nearly at the
free-streaming limit.  But the width of the shock as defined by the
temperature profile is roughly a mean-free path thick.  This is very
different than an inner, optically thick blackbody source assumed in
bare type Ia supernovae.  In addition, although the velocity profile
approaches a homologous outflow, it is not quite a homologous outflow.
Small variations in the shock temperature can cause large variations
in the emission, especially the UV emission.

This complexity, especially the fact that shock heating is playing a
role in the light-curve, makes it difficult to make detailed
predictions from enshrouded type Ia supernovae.  There are a few
robust claims.  First, enshrouded Ia supernova (with more than
0.1\,$M_\odot$ of surrounding material) should have broader V-band
light-curves.  The light-curves are, in part, powered by shock
heating, and these shocks lead to higher UV emission.  Thus far, the
current set of Swift UV type Ia supernova light-curves do not show
this high UV emission~\citep{Brown09}, suggesting that none of these
observed outbursts are enshrouded type Ia supernovae.

\begin{figure}
\epsscale{0.75}
\plotone{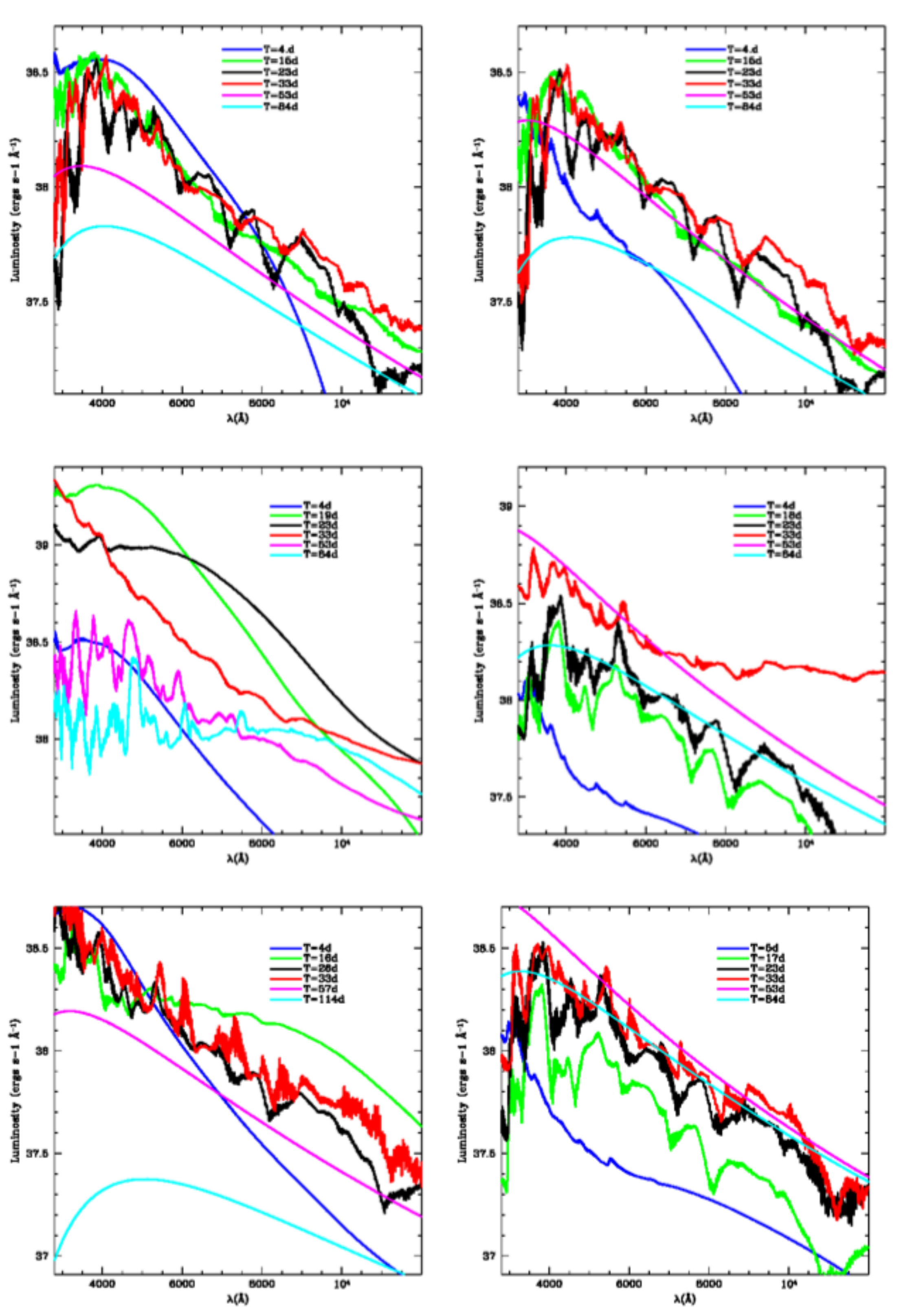}
\epsscale{1.0}
\caption{Spectra for our 6 primary models (See Fig.~\ref{fig:fullspec}
for details) focusing on the optical and infra-red spectra.}
\label{fig:specin}
\end{figure}

Strong UV emission and/or extremely broadened V-band peaks are
tell-tale signatures of massive C/O surroundings.
Figure~\ref{fig:specin} shows the full spectra for our models.  Near
peak, we see the strong calcium and silicon lines we expect in a type
Ia supernova.  With the uncertainties in the explosion and the 
outer density structure, it is possible to produce spectra at peak 
that fit many observed type Ia supernovae.  

But the spectra at early times are much more sensitive to the
surrounding medium.  For example, a broad absorption feature at $\sim
3700 \AA$ due to CaII H\&K and Si II is present at early times in
observed type Ia supernovae~\cite{Bufano09}, but it is not present
before peak in some of our models.  The surrounding C/O material
dominates the absorption features near peak, lessening the absorption
features by synthesized elements until peak.  Early observations may
be the key to finding distinguishing spectral features.

\begin{figure}
\epsscale{0.75}
\plotone{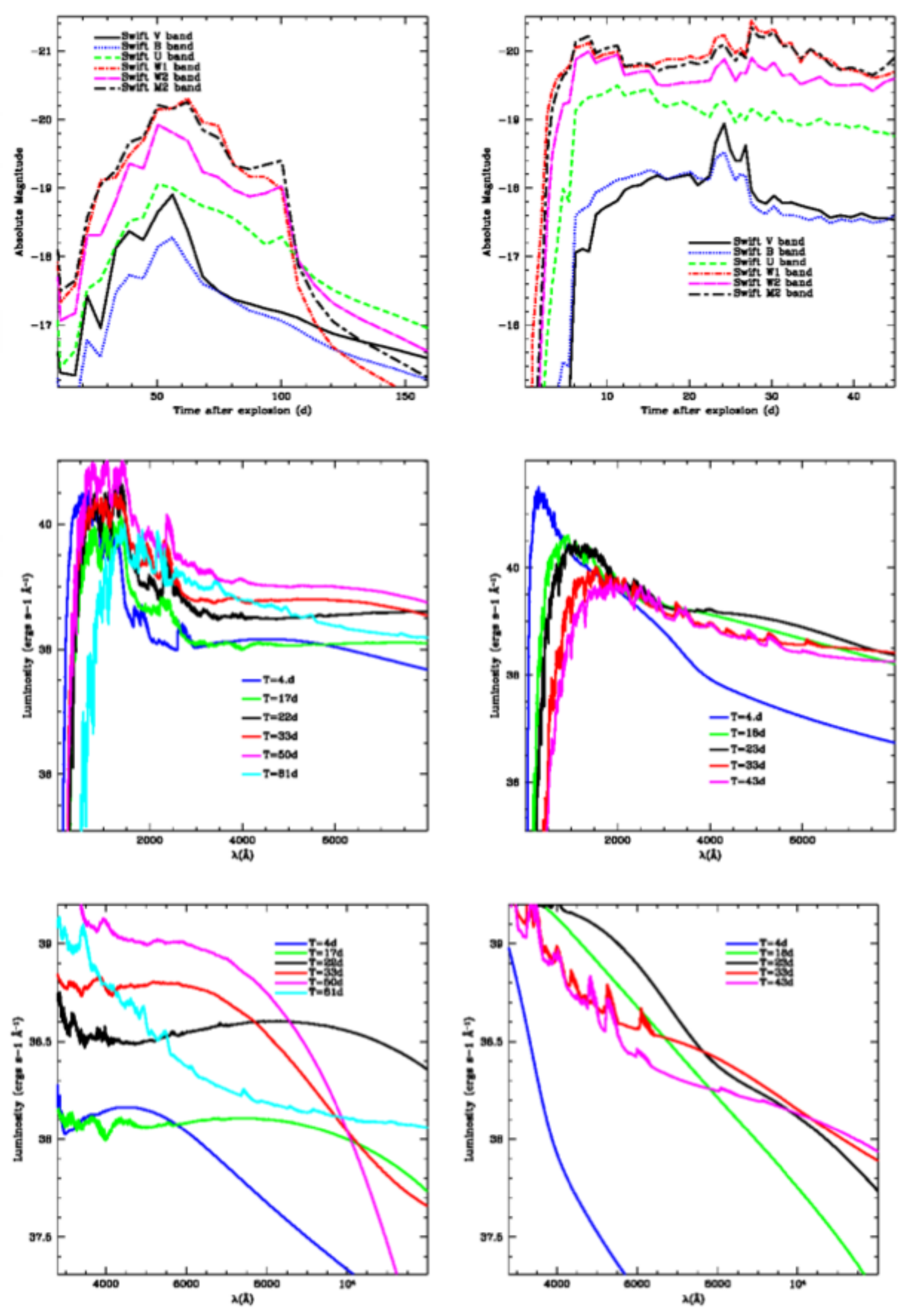}
\epsscale{1.0}
\caption{Light-curves (magnitude versus time) and spectra at specific 
times for 2 of our additional models:  0.35CO-Weak and 0.35CO-r3.  
Note that these models also have the long delays and enhanced UV 
emission.}
\label{fig:lcadd}
\end{figure}

We have run a series of other models, varying the explosion energy and
the density profile.  These models all show the same trends, extended
light-curves and enhanced UV emission.  For example,
figure~\ref{fig:lcadd} shows the light-curves and spectra of two of
these models: a weaker explosion and a shallower density profile.  
The peak in the weak explosion occurs later than our standard 
energy explosion, but the duration of the peak (e.g. the time 
the absolute visible magnitude is above -18) is roughly comparable 
to the standard explosion.  A shallower density profile produces 
a dimmer, but much longer duration, outburst.

We end with one cautionary note on exact line comparisons.  A major
difference in our models is that our post-process spectral
calculations are built on top of a true ``two-temperature'' radiation
hydrodynamics calculation.  By ``two-temperature'', we refer to the
fact that the material temperature is affected by shocks as well as
the radiation.  In typical type Ia supernova calculations, shock
heating is neglected (it is argued that does not play a role in the
observed emission for photons with wavelengths longer than the X-ray
or UV).  For these enshrouded progenitors, shock heating is more
important and our models include this effect.  But our models
currently lack the detailed resolution at the photosphere to catch
line features sensitive to the narrow region just beyond.  And we 
have not yet included the fact that the excitation levels of each 
atom are not described by a single equilibrium temperature.  Although 
the broad features discussed here will not change, many of the 
details of the spectra will change as more accurate models are made.

\section{Putting it all together}
\label{sec:summary}

In this paper, we have combined population synthesis studies of 
the double degenerate scenario for type Ia supernovae with 
merger calculations of these CO/CO binaries to produce the environments 
surrounding this type of type Ia supernova.  We have then modeled 
the propagation of a modern explosion simulation through this environment 
to produce light-curves of these enshrouded type Ia supernovae.  
The surrounding environment both delays and extends the X-ray flux 
in shock breakout, producing an initial X-ray signature much 
closer to the shock breakout of a type Ib/c supernova than that 
seen in the shock emergence of a bare type Ia.  The subsequent 
V-band peak is extended and is much broader than typical Ias.
Finally, although the peak and post-peak spectra will display the 
same lines as normal supernova, the early spectra will be dominated 
by carbon and oxygen lines only.

We have made a number of approximations in our calculations.  We
assumed that the density profile maintained its steep, $r^{-4}$
profile.  If the surrounding atmosphere redistributes, we would expect
a slightly flatter profile.  Our test model with a $r^{-3}$ profile
produced an initial X-ray flash that occurs later and the visible
light-curve is broader and dimmer.  Flattening the density profile 
produces an explosion very different than typical Ia supernovae.

Another approximation we made was to spherize the inherently
multi-dimensional merger and explosion models.  The merger debris is
asymmetric, slightly denser in the orbital plane.  The light curve
will be affected if the explosion is also asymmetric, sending a weaker
explosion along the orbital plane.  To test this affect, we modeled 
a weaker explosion.  Although the light-curve peaked later, it was 
similarly broad to our strong explosion, again not matching the 
typical type Ia supernova.

All of these results argue that these enshrouded systems can not
explain normal type Ia supernovae.  And, such systems must be less
common than normal type Ia supernovae to avoid dominating our current
Ia supernova sample.  Pakmor et al. (2010) argued that mergers, with a
weak explosion, might explain sub-luminous mergers.  They found that
the only issue with such an explanation is that the merger model
produced light-curves that were slightly broader than the
observations.  If they had run radiation-hydrodynamics calculations
(including the energy from shock heating), it is likely that the light
curves would be even broader, making an even worse fit to sub-luminous
models.

Astronomers have argued for other forms of enshrouded type Ia systems,
especially for the case of Supernova Ias with hydrogen envelopes.  For
example, supernova 2002ic is fit well by models invoking a wind
profile with mass loss rates in excess of 0.01\,M$_\odot \,{\rm
  y^{-1}}$ or circumstellar material extending out to
$10^{16}-10^{17}$\,cm with a possible gap between the exploding white
dwarf and this circumstellar
material~\citep{Chu02,Deng04,Kot04,Wood04}.  Supernova 2005gj may be
similar~\citep{Aldering06}.  Our surrounding debris is much more
compact, concentrated below $10^{12}$\,cm.

Some systems exist with strong carbon lines, suggesting incomplete 
C/O burning~\citep{mazzali01,Tan08}.  Typically these systems 
have considerable silicon from this partial burning.  If the 
envelopes of double degenerate mergers are particularly compact, 
we might be able to match these systems, but we haven't studied 
such systems here.

But if double degenerate systems are to dominate the Ia sample, why 
don't we observe them?

One explanation for the lack of observations of these enshrouded type
Ia supernovae is simply that CO/CO mergers do not produce type Ia
supernovae.  Nomoto and collaborators~\citep{NK91,NUK01} have argued that the
high accretion rate in these systems will preferentially lead to the
collapse of the merged white dwarf system to a neutron star.  The
resultant explosion will still produce a supernova-like explosion, but
could be much dimmer than what we studied here~\citep{Fry09}.  

However, other possible explanations exist.  Our current stellar
models predict too-large CO white dwarf masses.  There is some
evidence that the $^{12}$C/$^{12}$C fusion rate has been
under-estimated.  By increasing this rate, Herwig et al. (in
preparation) found that the maximum CO white dwarf mass is much lower
(as low as 0.8$M_\odot$).  In such a case, the maximum total merger
mass from CO binaries would be below 1.6$M_\odot$.  A number of ONe
binaries would exist in this scenario, but these mergers are expected
to collapse to neutron stars, yielding the dimmer supernova discussed
in the previous paragraph.

\begin{figure}
\plotone{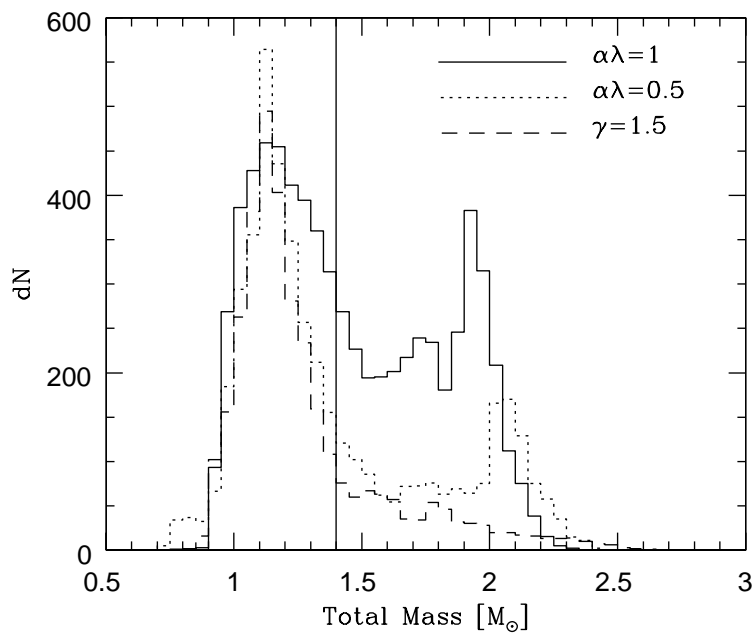}
\vspace{-1.8in}
\caption{Distribution of total masses for our population of merger
binaries based on the population synthesis models of Ruiter et
al. (2009).  We will assume only those binaries whose total 
mass exceeds 1.4\,M$_\odot$ will produce explosions.  Note 
that the peak in masses above 1.4\,M$_\odot$ occurs at 
roughly 2\,M$_\odot$.}
\label{fig:popsynth}
\end{figure}

Another potential explanation is that our standard population
synthesis models predict a peak at around 2M$_\odot$ in the total mass
distribution of type-Ia producing CO/CO mergers, arguing for massive
surrounding envelopes ($\sim 0.6M_\odot$).  But by modifying the
prescription for binary mass transfer, i.e., by adopting the
$\gamma$-formalism (model 3) rather than the $\alpha$-formalism, the
peak at $~2\,M_\odot$ disappears, arguing that most mergers with
masses above the Chandrasekhar limit occur with masses within
0.1\,M$_\odot$ of this limit~\ref{fig:popsynth}.  Without a better
understanding of common envelope evolution, population synthesis
models may simply be unable to predict the mass distribution of merger
WD binaries accurately.  Perhaps there are simply few enshrouded type
Ias and most systems only have $0.1\,M_\odot$ of surrounding material
or less.  If such systems can explode to form type Ia supernovae,
their observations would not be so distinct from the current type Ia
population.

What we can say for certain is that the spectral and light-curve
features of double degenerate mergers with total system masses in
excess of roughly 1.5-1.6$M_\odot$ do not match normal type Ia
supernovae.  These super-Chandrasekhar mergers can not dominate 
Ia progenitors.

\acknowledgements This work was carried out in part under the auspices
of the National Nuclear Security Administration of the U.S. Department
of Energy at Los Alamos National Laboratory and supported by Contract
No. DE-AC52-06NA25396.

\clearpage

\begin{deluxetable}{lcccccc}
\tablewidth{0pt}
\tablecaption{Transient Models}
\tablehead{
  \colhead{Name}
& \colhead{M$_{\rm Envelope}$}
& \colhead{Envelope}
& \colhead{V$_{\rm peak}$} 
& \colhead{V$_{\rm FWHM}$}
& \colhead{UV$_{\rm peak}$}
& \colhead{UV}\\
  \colhead{}
& \colhead{M$_\odot$}
& \colhead{Composition}
& \colhead{Abs. Mag.}
& \colhead{d}
& \colhead{Abs. Mag.}
& \colhead{Features}

}

\startdata

0.1co & 0.1 & CO & -17 & 35 & -17 & Double Peak \\
0.1he & 0.1 & He & -17 & 45 &  -17 & Double Peak \\
0.35co & 0.35 & CO & -19 & 100 & -20 & Double Peak \\
0.35he & 0.35 & He & -17 & 30 &  -20 & Broad (40\,d) \\
0.7co & 0.7 & CO & -17 & 100 & -18 & Delayed (40\,d rise) \\
0.7he & 0.7 & He & -17 & 100 &  -18 & Broad ($> 40$\,d) \\
0.35co Weak & 0.35 & CO & -19 & 40 & -20 & Delayed (60\,d rise) \\ 
0.1co Weak & 0.1 & CO & -17 & 40 & -17 & Delayed (100\,d rise) \\ 
0.35co-r3 & 0.35 & CO & -19 & $>100$ & -20 & Broad \\

\enddata

\label{tab:sims}
\end{deluxetable}

\clearpage

\clearpage

\clearpage

\clearpage

\clearpage

\clearpage

\clearpage

\clearpage

\clearpage

\clearpage

\clearpage

\clearpage

\clearpage


\begin{thebibliography}

\bibitem[Aldering et al.(2006)]{Aldering06} Aldering, et al. 2006, 
ApJ, 650, 510

\bibitem[Belczynski, Kalogera \& Bulik(2002)]{Bel02} Belczynski, 
K., Kalogera, V., \& Bulik, T. 2002, ApJ, 572, 407

\bibitem[Belczynski et al.(2008)]{Bel08} Belczynski, 
K., Kalogera, V., Rasio, F.A., Taam, R.E., Zezas, A., Bulik, T., 
Maccarone, T.J., \& Ivanova, N. 2008, ApJS, 174, 223

\bibitem[Brown et al.(2009)]{Brown09} Brown, P.J., Holland, S.T., 
Immler, S., Milne, P., Roming, P.W.A., Gehrels, N., Nousek, J., Panagia, N., 
Still, M., Vanden Berk, D. 2009, AJ, 137, 4517 

\bibitem[Bufano et al.(2009)]{Bufano09} Bufano, F., Immler, S., Turatto, M., 
Landsman, W., Brown, P., Benetti, S., Cappellaro, E., Holland, S.T., Mazzali, P., 
Milne, P., Panagia, N., Pian, E., Roming, P., Zampieri, L., Breeveld, A.A. 
Gehrels, N. 2009, ApJ, 700, 1456

\bibitem[Colgate et al.(1993)]{Col93}
Colgate, S.A., Herant, M., \& Benz, W. 1993, Phys. Rep., 227, 157

\bibitem[Chugai et al.(2004)]{Chu02} Chugai, N.N., Chevalier, R.A., 
Lundqvist, P. 2004, MNRAS, 355, 627

\bibitem[Deng et al.(2004)]{Deng04} Deng, J., Kawabata, K.S., Ohyama, Y., 
Nomoto, K., Mazzali, P.~A., Wang, L., Jeffery, D.J., Iye, M., Tomita, H., 
\& Yoshii, Y. 2004, 605, L37

\bibitem[de Kool(1990)]{Kool90} de Kool, M., 1990, ApJ, 358, 189

\bibitem[Fryxell et al.(2000)]{Fryx00}
Fryxell, B., Olson, K., Ricker, P., Timmes, F.X., Zingale, M., Lamb, D.Q., 
MacNeice, P., Rosner, R., Truran, J.W., Tufo, H. 2000, ApJS, 131, 273

\bibitem[Fryer et al.(2006)]{Fry06} Fryer, C.L., Rockefeller, G., 
\& Warren, M.S. 2006, ApJ, 643, 292

\bibitem[Fryer et al.(2007)]{Fry07} Fryer, C.L., Hungerford, A.L., 
\& Young, P.A. 2007, ApJ, 662, L55

\bibitem[Fryer et al.(2009)]{Fry09} Fryer et al. 2009, ApJ, 707, 193

\bibitem[Gittings et al.(2008)]{Git08}
Gittings, M., Weaver, R., Clover, M., Betlach, T., Byrne, N., 
Coker, R., Dendy, E., Hueckstaedt, R., New, K., Oakes, W.R., 
Ranta, D., Stefan, R. 2008, Comp. Sci. and Disc. 1, 015005

\bibitem[H\"oflich(2005)]{Hof05} H\"oflich, P. 2005, in 1604-2004:
Supernovae as Cosmological Lighthouses, ASP Conference Series,
Vol. 342, Proceedings of the conference held 15-19 June, 2004 in
Padua, Italy. Edited by M. Turatto, S. Benetti, L. Zampieri, and
W. Shea. San Francisco: Astronomical Society of the Pacific, 2005.,
p.372

\bibitem[H\"oflich \& Schaefer(2009)]{Hof09} H\"oflich, P. \& Schaefer, 
B.E. 2009, ApJ, 705, 483

\bibitem[Houck \& Chevalier(1992)]{Hou92}
Houck, J.C. \& Chevalier, R.A. 1992, ApJ, 395, 592

\bibitem[Kotak et al.(2004)]{Kot04} Kotak, R., Meikle, W.P.S., 
Adamson, A., Leggett, S.K. 2004, MNRAS, 354, L13

\bibitem[Livio(2001)]{Liv01} Livio, M. 2001, in Supernovae and
Gamma-Ray Bursts: The Greatest Explosions since the Big Bang,
ed. M. Livio, N. Panagia, \& K. Sahu (Cambridge: Cambridge
Univ. Press), 334 

\bibitem[Magee et al.(1995)]{Mag95}
Magee, N.H., Abdallah, J. Jr., Clark, R.E.H., Cohen, J.S., 
Collins, L.A., Csanak, G., Fontes, C.J., Gauger, A., Keady, J.J., 
Kilcrease, D.P., Merts, A.L. 1995, Astron. Soc. of the Pac., 78, 51

\bibitem[Mazzali et al.(2001)]{mazzali01}
Mazzali, P.~A. 2001, MNRAS, 321, 341

\bibitem[Mazzali et al.(2007)]{Maz07}
Mazzali, P.~A., R{\"o}pke, F.~K., Benetti, S., \& 
Hillebrandt, W. 2007, Science, 315, 825

\bibitem[Meakin et al.(2009)]{Mea09}
Meakin, C.A., Seitenzahl, I., Townsely, D., Jordan, G.C., 
Truran, J., Lamb, D. 2009, ApJ, 693, 1188

\bibitem[Nelemans et al.(2000)]{Nel00} Nelemans, G., 
Verbunt, F., Yungelson, L.R., Portegies-Zwart, S.F.
2000, A\&A, 360, 1011

\bibitem[Nelemans \& Tout(2005)]{Nel05} Nelemans, G., \& 
Tout, C.A. 2005, MNRAS, 356, 753

\bibitem[Nomoto \& Kondo(1991)]{NK91}
Nomoto, K., \& Kondo, Y. 1991, ApJ, 367, L19

\bibitem[Nomoto et al.(2001)]{NUK01}
Nomoto, K., Umeda, H., \& Kobayashi, C. 2001, in Birth and Evolution 
of the Universe, proceedings of the 4th RESCEU International Symposium, 
ed. K. Sato \& M. Kawasaki (Universal Academy Press), 235

\bibitem[Pakmor et al.(2010)]{Pak10}
Pakmor, R., Kromer, M., Roepke, F.~K., Sim, S.~A., Ruiter, A.~J., 
\& Hillebrandt, W. 2009, Nature, 463, 61

\bibitem[Piro et al.(2010)]{Pir10}
Piro, A.L., Chang, P., \& Weinberg, N.N. 2010, ApJ, 708, 598

\bibitem[Plewa et al.(2004)]{Ple04}
Plewa, T., Calder, A.C., \& Lamb, D.Q. 2004, ApJ, 612, L367

\bibitem[Raskin et al.(2009)]{Rask09} Raskin, C., 
Timmes, F.~X., Scannapieco, E., Diehl, S., \& Fryer, C. 2009, 
\mnras, 399, L156

\bibitem[Rosswog et al.(2009)]{Ross09} Rosswog, S., Kasen, D., 
Guillochon, J., \& Ramirez-Ruiz, E. 2009, \apjl, 705, L128

\bibitem[Ruiter, Belczynski \& Fryer(2009)]{Rui09} Ruiter, A.J,
Belczynski, K., \& Fryer, C. 2009, ApJ, 699, 2026

\bibitem[Tanaka et al.(2008)]{Tan08} Tanaka, M., Mazzali, P.A., 
Benetti, S., Nomoto, Ken'ichi, Elias-Rosa, N., Kotak, R., Pignata, G., 
Stanishev, V., Hachinger, S. 2008, ApJ, 677, 448

\bibitem[Webbink(1984)]{Web84} Webbink, R.F. 1984, ApJ, 277, 355

\bibitem[Wood-Vasey et al.(2004)]{Wood04} Wood-Vasey, W.M., Wang, L., 
\& Aldering, G. 2004, 616, 339

\bibitem[Yoon et al.(2007)]{Yoon07}
Yoon, S.-C., Podsiadlowski, P., \& Rosswog, S. 2007, \mnras, 380, 933

\end{thebibliography}
\end{document}